\documentclass{paper}

\usepackage{fullpage}

\usepackage{graphicx}

\usepackage[font=footnotesize]{subcaption}

\usepackage{csquotes}

\usepackage{xspace}
\usepackage{siunitx}
\usepackage[sort&compress, numbers]{natbib}

\usepackage{acronym}
\acrodef{udel}[UDel]{the University of Delaware}

\usepackage{tabularx, booktabs}

\usepackage[inline]{enumitem}
\setlist[enumerate]{label=(\arabic*), itemjoin={,\ }, itemjoin*={,\ and\ }}

\usepackage{hyphenat}
\usepackage{authblk}

\usepackage{amsmath}
\DeclareMathOperator{\rank}{rank}

\usepackage[capitalize,sort]{cleveref} 
\crefname{figure}{Figure}{Figures}
\crefname{task}{Task}{Tasks}

\author[1]{Jianwei Wu}
\author[2]{Jining Yu}
\author[3]{David Shepherd}
\author[4]{James Clause}
\affil[1,2,4]{University of Delaware}
\affil[3]{Virginia Commonwealth University}

\title{Shortcomings of Class-level Documentation: A Survey }

\begin{document}

\maketitle

To better understand the shortcomings of class-level documentation we conducted a survey of \num{167} experienced software developers recruited from from ABB, Inc., \ac{udel}, and popular, on-line programming forums (e.g., the programming subreddit\footnote{https://www.reddit.com/r/programming/}).
As part of this survey, we asked participants two questions: \enquote{Q1: How frequently do you read class-level documentation because you want to learn: \emph{information need}?} and \enquote{Q2: How frequently does class-level documentation sufficiently explain: \emph{information need}?}.
Responses to both of these questions were collected using a 3-point Likert scale: Rarely, Sometimes, Often.
The specific information needs we asked about were selected based on our experience and knowledge of class-level documentation as well as investigations into questions developers ask during software development (e.g.,~\cite{sillito2006questions, Campos2014, Barua2014, LaToza2010, maalej:2013}.)

\begin{figure*}[t]
  \centering
  \includegraphics[width=\textwidth]{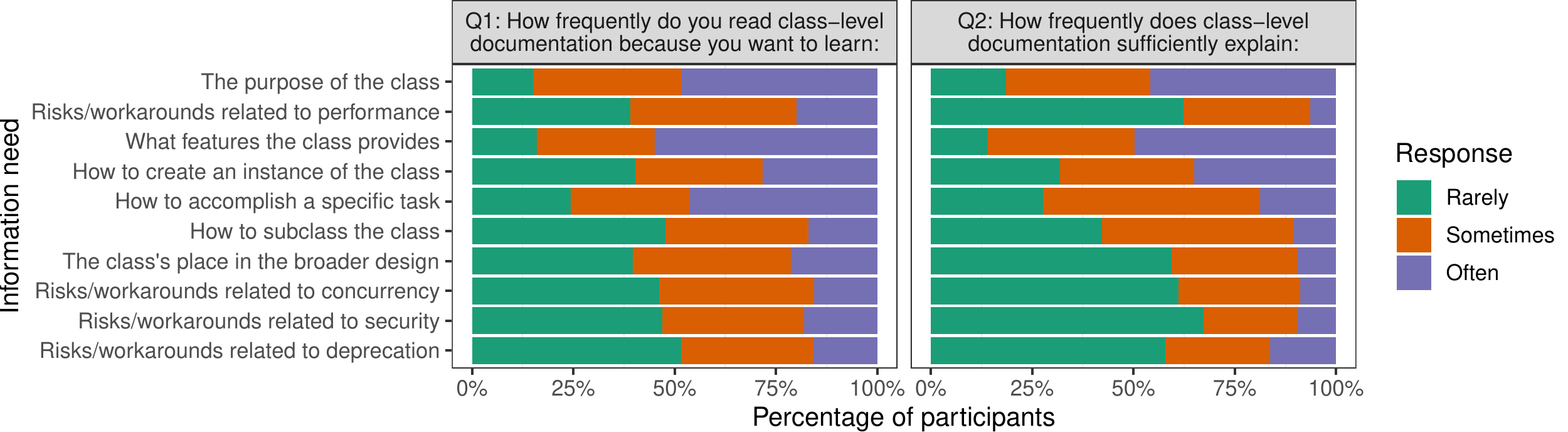}
  \caption{Distribution of participant responses to \enquote{Q1: How frequently do you read class-level documentation because you want to learn:} (left) and \enquote{Q2: How frequently does class-level documentation sufficiently explain:} (right) for each information need.}
  \label{fig:responses}
\end{figure*}

\Cref{fig:responses} presents two stacked bar charts that show the distribution of participant responses to our survey questions.
The left-hand chart shows the responses to Q1 and the right-hand chart shows the responses to Q2.
In each chart, the y-axis shows the information needs that were asked about and the stacked bars show the proportion of participants (x-axis) who answered Rarely (left, green), Sometimes (middle, orange), and Often (right, purple).
For example, the top-most stacked bar in the left-hand chart shows that when asked \enquote{How frequently does class-level documentation sufficiently explain: The purpose of the class}, \SI{\approx 18}{\percent} of participants answered Rarely, \SI{\approx 36}{\percent} answered Sometimes, and \SI{\approx 46}{\percent} answered Often.

To help us understand participants' views on the quality of class-level documentation and to identify opportunities for improvement, we calculated each participant’s annoyance level with respect to each information need.
Intuitively, a participant's annoyance level is a relative approximation of how frequently the participant is frustrated when attempting to satisfy an information need that takes into account both how frequently they have the information need (Q1) and how frequently class-level documentation sufficiently explains the information need (Q2).
For example, if a participant believes that class-level documentation Often satisfies an information need they are unlikely to be frustrated, regardless of how often they want to learn about the need, and their annoyance level should be negligible.
Conversely, if a participant believes that class-level documentation Rarely or Sometimes satisfies an information need, they are likely to be frustrated in proportion to how frequently they want to learn about the need, with the highest level of annoyance occurring when they Often want to learn about a need that is Rarely satisfied.

To formalize the intuitive definition of annoyance described above, we define a participant's annoyance level \(a\) with respect to an information need \(need\) as follows:
\[
a_{need} =
\begin{cases}
    0,                                       & \text{if } Q2_{need} \text{ is Often} \\
    \rank(Q1_{need}) - \rank(Q2_{need}) + 2, & \text{otherwise}
\end{cases}
\]
where \(Q1_{need}\) and \(Q2_{need}\) are the participant's responses to Q1 and Q2 when asked about \(need\), respectively, and \(\rank\) returns the rank of the response where \(\text{Often} > \text{Sometimes} > \text{Rarely}\).
For example, if, when asked about \(need\), a participant responded Rarely to Q1 (\(\rank(Q1_{need}) = 1\)) and Sometimes to Q2 (\(\rank(Q2_{need}) = 2\)), their annoyance level would be \num{1} (\(a_{need} = 1 - 2 + 2 = 1\)).
Alternatively, if they responded Sometimes to Q1 (\(\rank(Q1_{need}) = 2\)) and Rarely to Q2 (\(\rank(Q2_{need}) = 1\)), their annoyance level would be \num{3} (\(a_{need} = 2 - 1 + 2 = 3\)).
The output of this equation ranges from \numrange{0}{4}, with \num{0} indicating that the participant is likely not annoyed by the information need (i.e., when a participant believes that class-level documentation Often satisfies an information need, regardless of how frequently they want to know about the need) and numbers greater than zero indicating increasing levels of annoyance with \num{4} as the highest level of annoyance (i.e., when a participant Often wants to learn about an information need that class-level documentation Rarely sufficiently explains).

\begin{figure*}[t]
  \centering
  \includegraphics[width=\textwidth]{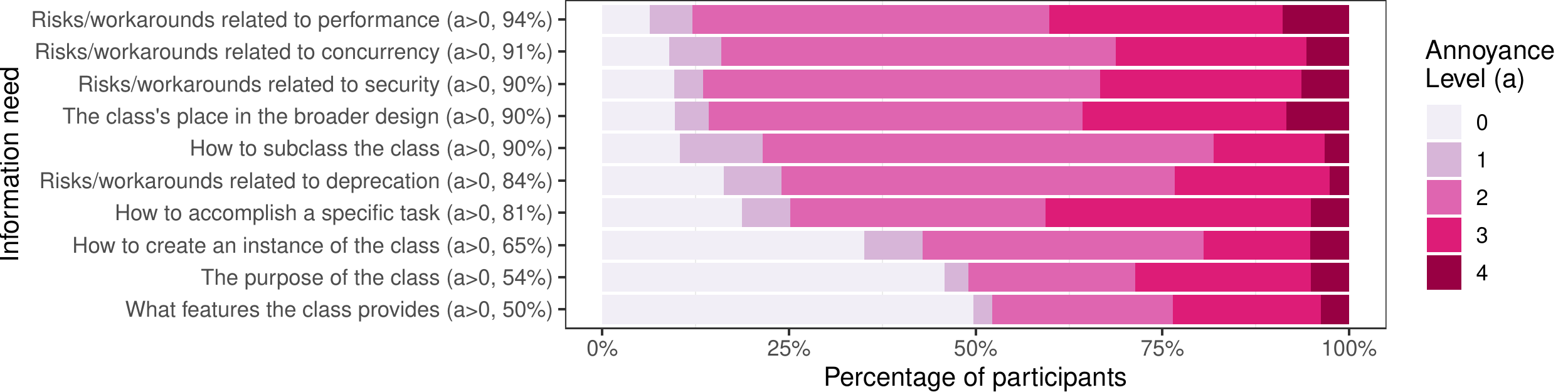}
  \caption{Distribution of participant annoyance levels for each information need sorted by proportion of participants with non-zero annoyance levels (\(a > 0\)) in decreasing order from top to bottom.}
  \label{fig:annoyance}
\end{figure*}

\Cref{fig:annoyance} shows the distribution of annoyance levels for our participants as a stacked bar chart.
As in \cref{fig:responses}, the y-axis shows the information needs that were asked about and the stacked bars show the proportion of participants at each annoyance level (x-axis).
To simplify interpretation, annoyance levels are sorted in increasing order from left to right with the least annoyance level colored gray and higher-levels of annoyance shown in increasingly darker shades of red.
For example, the top stacked bar shows that for Risks\slash workarounds related to performance, \SI{\approx 9}{\percent} of respondents have an annoyance level of \num{4}, \SI{\approx 31}{\percent} of respondents have an annoyance level of \num{3}, \SI{\approx48}{\percent} of respondents have an annoyance level of \num{2}, and \SI{\approx 6}{\percent} of respondents have an annoyance level of \num{1}, while only \SI{\approx 6}{\percent} of respondents have an annoyance level of \num{0}.
To simplify comparisons between the overall levels of annoyance for each information need, the information needs are sorted in decreasing order based on the percentage of participants with a non-zero annoyance level.
As a result, the information need that the largest proportion of respondents were annoyed with to some degree (Risks\slash workarounds related to performance, \(a>0\), \SI{\approx 94}{\percent}) is shown at the top of the figure and the information need that the smallest proportion of respondents were annoyed with to some degree (What features the class provides \(a>0\), \SI{\approx 50}{\percent}) is shown at the bottom.

The predominance of red this figure clearly shows that developers are annoyed with class-level documentation.
For example, there are five information needs where more than \SI{90}{\percent} of participants have a non-zero level of annoyance.
Moreover, even in the case of the least annoying information need (What features the class provides), \SI{50}{\percent} of participants have a non-zero annoyance level.
This result strongly motivates our proposed work by demonstrating
\begin{enumerate*}
  \item the importance of class-level documentation to developers as a potential resource for satisfying information needs (i.e., developers often consult class-level documentation)
  \item the inability of class-level documentation to regularly satisfy such information needs
\end{enumerate*}.

\bibliographystyle{unsrtnat}
\bibliography{paper}

\end{document}